Fractionation by shape in deterministic lateral displacement microfluidic devices

Mingliang Jiang, Kostyantyn Budzan and German Drazer

Mechanical and Aerospace Engineering Department, Rutgers, The State University of New Jersey, Piscataway, NJ.

**Abstract**

We investigate the migration of particles of different geometrical shapes and sizes in a scaled-up model of a gravity-driven deterministic lateral displacement (g-DLD) device. Specifically, particles move through a square array of cylindrical posts as they settle under the action of gravity. We performed experiments that cover a broad range of orientations of the driving force (gravity) with respect to the columns (or rows) in the square array of posts. We observe that as the forcing angle increases particles initially locked to move parallel to the columns in the array begin to move across the columns of obstacles and migrate at angles different from zero. We measure the probability that a particle would move across a column of obstacles, and define the critical angle $\theta_c$ as the forcing angle at which this probability is 1/2. We show that critical angle depends both on particle size and shape, thus enabling both size- and shape-based separations. Finally, we show that using the diameter of the inscribed sphere as the characteristic size of the particles the corresponding critical angle becomes independent of particle shape and the relationship between them is linear. This linear and possibly universal behavior of the critical angle as a function of the diameter of the inscribed sphere could provide guidance in the design and optimization of g-DLD devices used for shape-based separation.

**Introduction**

The ability to use microfluidics systems to fractionate a mixture of suspended particles based on their shape could lead to important applications. In the case of bioparticles, for example, shape is an important factor used to identify and characterize them. In fact, cell morphology has long been used in clinical diagnosis and it is also a marker of cell cycle.[1–3] At the nanoscale, nanoparticles exhibit shape-dependent properties,[4] including mechanical,[5] optical[6] and catalytic properties.[7] In addition, nanoparticles are sometimes the building blocks of larger assemblies, and shape could determine the emerging structures.[8] At even smaller, molecular scales, the potential separation of enantiomers using microdevices was recently investigated.[9]

A number of different methods have recently been investigated to separate microparticles and nanoparticles by shape.[1–3,10–13] In microfluidics, a promising separation method is deterministic lateral displacement (DLD), which offers continuous two-dimensional separation of suspended particles.[14–17] DLD has been particularly successful in the separation of biological samples.[2,18–31] However, only a few studies have investigated the role of shape in DLD separation methods. Tegenfeldt and coworkers investigated the motion of red blood cells in devices of different heights, greater and smaller than the largest dimension of the cells.[2] In this way, they were able to control cell orientation and showed that trajectories observed in the case of narrow channels are different from those observed when the cells are free to take any orientation. Also working with red blood cells, Zhang and coworkers explored the use of non-cylindrical posts to enhance separation.[32] In these two cases only the shape of red blood cells was considered. Reinman and coworkers demonstrated the separation of (planar) chiral particles, that is the separation between so-called "L"- and "T"-shaped particles restricted to move in the plane of the device.[9]

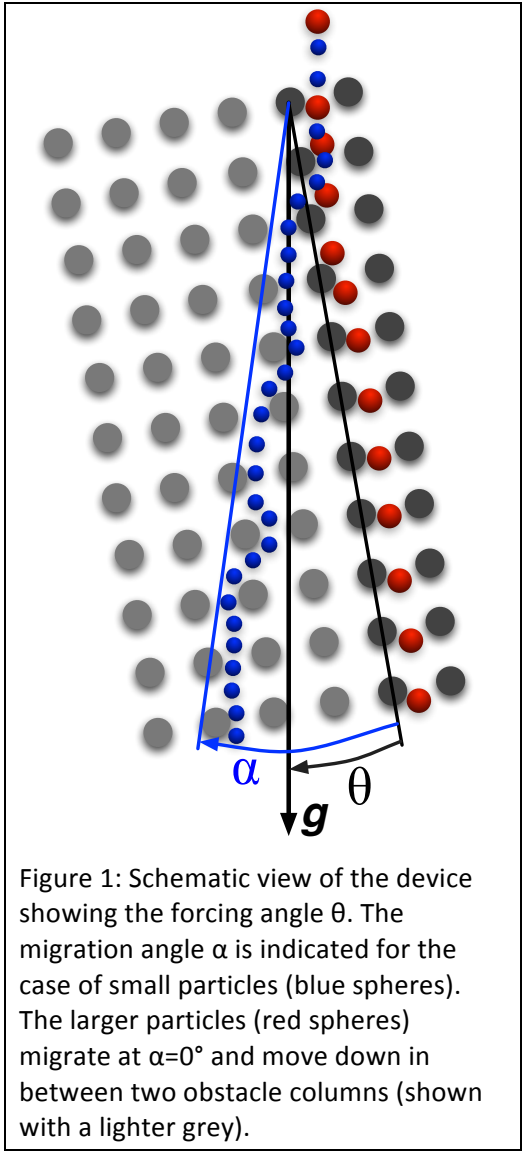

Figure 1: Schematic view of the device showing the forcing angle θ. The migration angle α is indicated for the case of small particles (blue spheres). The larger particles (red spheres) migrate at α=0° and move down in between two obstacle columns (shown with a lighter grey).

Here, we investigate the trajectory followed by particles of different shape to identify conditions leading to separation. In previous experimental and analytical work using spherical particles, we have extended the method to gravity-driven DLD (g-DLD) and showed the presence of a critical forcing angle.[33,34,15,35–37] The forcing angle, θ, is defined by the direction of gravity with respect to the array of obstacles and the migration angle is defined by the average orientation of the particles trajectory also with respect to the array (see Figure 1 for a schematic view). For small enough angles, particles move along a column of obstacles in the device, as shown for one of the particles in Figure 1. The critical angle, $\theta_c$, is defined as the forcing angle at which the particles move across columns of obstacles and migrate at an angle different from zero. In the case of non-spherical particles, not all the particles of a given species will cross at the same angle and therefore we calculate the probability $p_\beta$ that a particle of species $\beta$ would cross a column of obstacles (or *crossing probability*) and define the critical angle $\theta_c$ for species $\beta$ as the forcing angle at which $p_\beta = 1/2$. It is clear, that particles with different critical angles can be separated at intermediate values of the forcing angle. For example, in the case represented in Figure 1 the critical angle of the smaller particles is smaller than the forcing angle, whereas the critical angle of the larger particles is larger, and the suspension could be easily fractionated. In fact, in previous experiments, we have consistently found that the critical angle increases with particle size.

**Experimental Set-up**

We have performed our experiments using macroscopic systems. Specifically, we scaled up a DLD device from micrometer to millimeter scale. Using such macroscopic systems simplifies several aspects of the experimental work significantly, including capturing the details of particle trajectories, re-using the same experimental system in all the experiments and covering a range of forcing angles with a single device. In previous studies we have maintained a small Reynolds number to compare the results directly with those obtained in microdevices. Here, we simplify the system even further using water as the fluid and, as a result, performing experiments at larger

Reynolds numbers. In particular, the Reynolds numbers based on particle velocity are O(100) or smaller. These Reynolds numbers are, therefore, within reach of microdevices working at high throughput.

A schematic view of the separation system is shown in Figure 2. It is composed of three main parts: the lattice board that houses the obstacle lattice and the rotating mount. The *obstacle lattice* was fabricated with a 3D printer (Objet350 Connex, Stratasys) and is 200mm x 200mm with 35 x 35 cylindrical posts. The diameter of the posts, or *obstacles,* is 2R=1mm and the spacing between two adjacent obstacles is 5mm. The obstacle lattice goes inside the recess in the *lattice board* and is covered by a piece of acrylic glass, thus creating the *separation system*. Then, the system is filled with water and lined up vertically so that the gravity force is on the plane of the obstacle array. A valve was set at the bottom of the system and it is used to drain water and collect particles between experiments. The separation device is then mounted on a rotating disk (*rotating mount),* which allows adjusting the orientation of the driving force with respect to the obstacle lattice.

We also used the 3D printer to fabricate particles of different shapes: cubes, cylinders, pyramids, spheres and tetrahedrons, as shown in Figure 2. Their density is 1.13 g/cm$^3$. We use a characteristic dimension of the particles as their *nominal size a.* Specifically, the nominal size of cubes, pyramids and tetrahedrons is the length of their edges (or sides of each facet), that of spheres is their diameter and for cylinders both the diameter and height. Following previous work, which indicates that larger aspect ratios (*a/R>1*) leads to better resolution, we used particles that are larger than the obstacles.[38] In addition, we noticed that smaller particles whose diameter is less than that of the cylindrical posts could end up finding equilibrium positions on top of the posts and would stop moving. On the other hand, the largest particles that can be separated are determined by the spacing between the obstacles. Based on these considerations, we performed experiments with particles ranging from 1.5 mm (smallest cubes) to 3.5 mm (largest tetrahedrons).

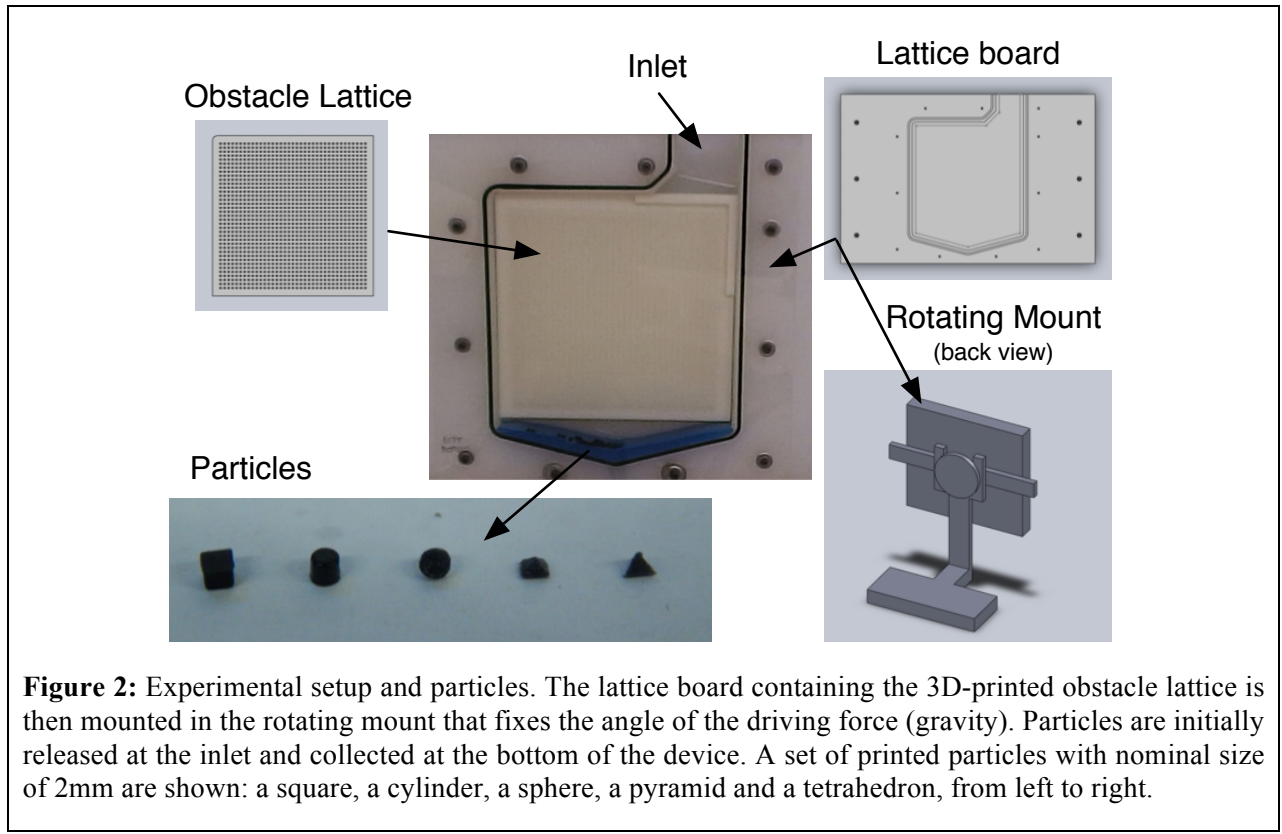

**Figure 2:** Experimental setup and particles. The lattice board containing the 3D-printed obstacle lattice is then mounted in the rotating mount that fixes the angle of the driving force (gravity). Particles are initially released at the inlet and collected at the bottom of the device. A set of printed particles with nominal size of 2mm are shown: a square, a cylinder, a sphere, a pyramid and a tetrahedron, from left to right.

In all the experiments, the particles are released from the top of the tank and they settle through the array of cylindrical obstacles. In each experiment the angle is controlled with the rotating mount. For each particle type we record the motion of one hundred particles and get their trajectories using ImageJ. The crossing probability of a given species is then evaluated as the number of particles moving across a column of obstacles over the total number of particles. In this calculation we only consider particles that move through the entire device (some particles, especially at forcing angles close to the critical one, stop moving and settle on top of an obstacle).

**Results and discussion**

In Figure 3 we present the summary of our results, investigating the motion of particles of five different shapes and four different sizes each for a range of angles around the respective critical angles. In all cases we plot the crossing probability $p_\beta$ of species $\beta$ as a function of the forcing angle $\theta$. In agreement with previous work with spherical particles, there is a range of driving angles different from zero for which the probability of crossing is zero. We usually refer to these cases in which the particles migrate at $\alpha = 0°$ when $\theta \neq 0°$ as *directional locking*.[15,16,34] At larger angles, the crossing probability displays a clear transition from no-crossing $p_\beta = 0$ to complete crossing $p_\beta = 1$ over a relatively narrow range of driving angles. As we discussed before, we define the critical angle $\theta_c$ as the angle at which the crossing probability is $p_\beta = 1/2$. In general, the critical angle increases with size, which would allow for size-based separation of non-spherical particles of different geometries. The only case in which the resolution is not enough for size-based separation is that of pyramids, which show almost identical probability curves independent of the sizes considered here (see Figure 3e).

More importantly, the results show that it is possible to separate particles of the same nominal size but significantly different shape. In order to represent this more explicitly, in Figure 3f we present the critical angle for the different types of particles as a function of their nominal size. It is clear that, although spheres, cubes and cylinders share similar critical angles for the same nominal size, their critical angles are different from those of pyramids and tetrahedrons, thus making it possible to separate between them. For example, at a forcing angle $\theta = 15°$ most of the spheres, cubes and cylinders of 2mm would not cross a column of obstacles and thus migrate at $\alpha = 0°$ displaying directional locking. On the other hand, the probability of crossing for tetrahedrons would be unity and that of pyramids would be close to 75%, and they would not be directionally locked, with an average migration angle $\alpha > 0°$ in both cases.

These results, however, do not directly show the difference in migration angles between locked and unlocked species or the associated resolution of the separation. Therefore, we also measured the migration angles directly, for forcing angles close to the critical angle in each case. The average migration angles are presented in Figure 4 for 2mm spheres, cubes, cylinders and tetrahedrons.

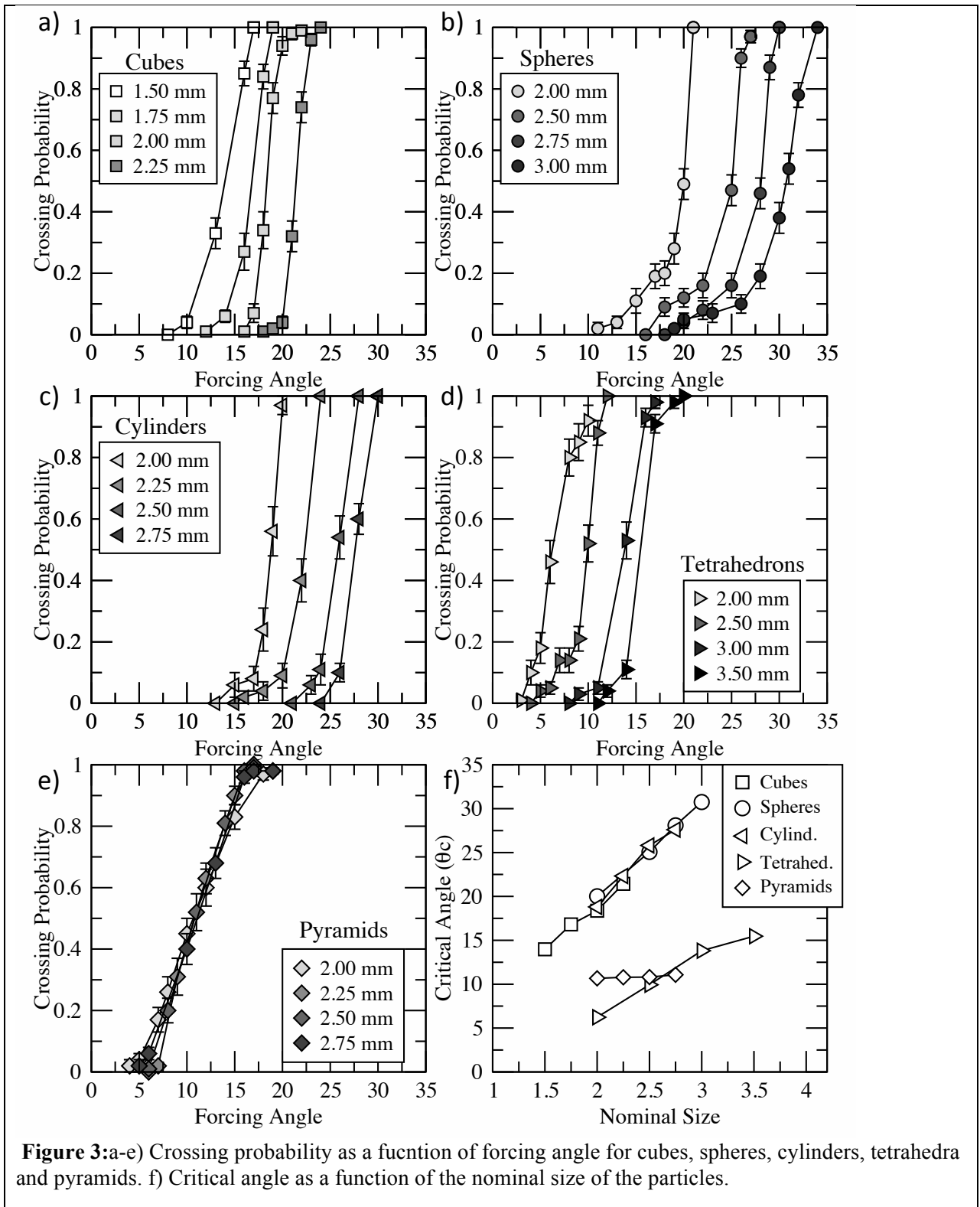

**Figure 3:** a-e) Crossing probability as a fucntion of forcing angle for cubes, spheres, cylinders, tetrahedra and pyramids. f) Critical angle as a function of the nominal size of the particles.

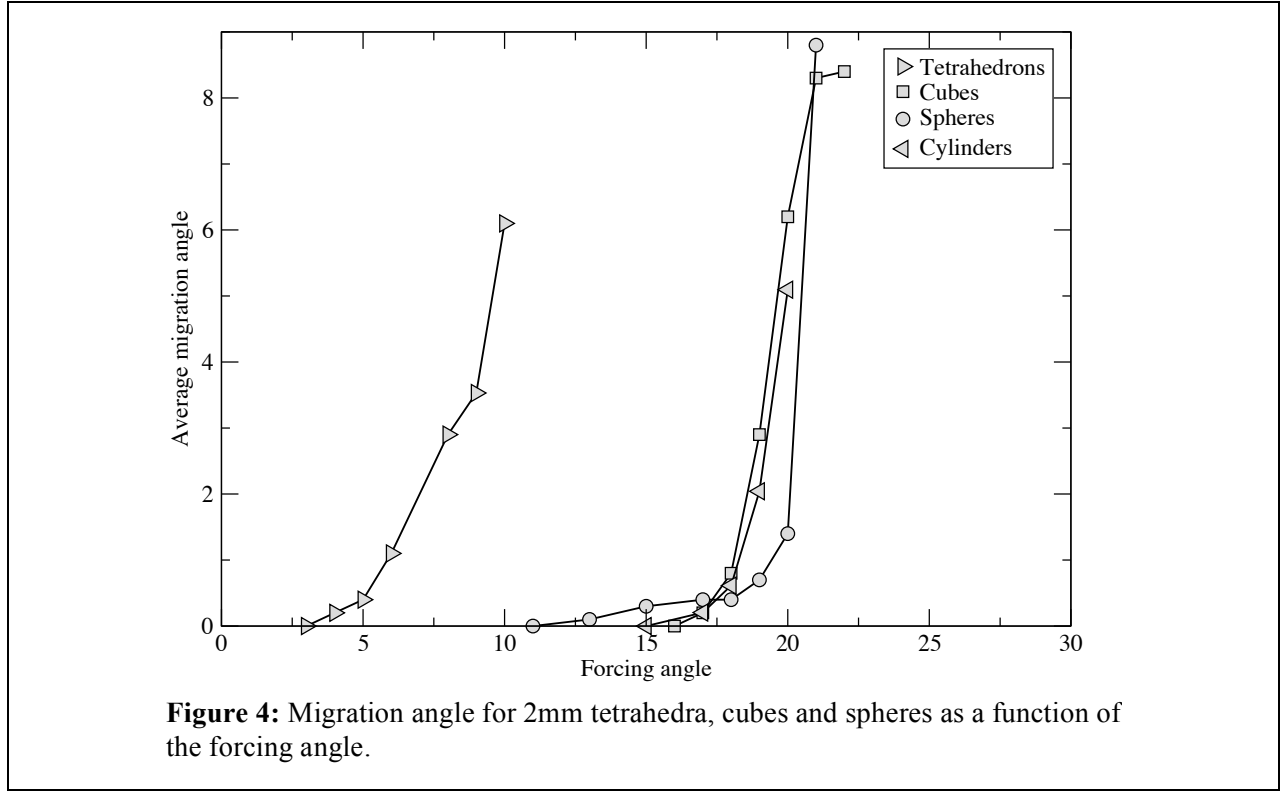

**Figure 4:** Migration angle for 2mm tetrahedra, cubes and spheres as a function of the forcing angle.

As expected, the largest difference is between the tetrahedrons migrating at angles around $\alpha \cong 6°$ for $\theta \cong 10°$ and the rest of the particles (cubes, cylinders and spheres) that are still locked to move at $\alpha = 0°$. In order to quantify the quality of the separation, we also measured the entire probability distribution (or histogram) of migration angles. In Figure 5 we present the probability distribution for cubes for the forcing angles presented in Figure 4, corresponding to the range of forcing angles over which the transition from no-crossing ($p_\beta = 0$) to complete crossing ($p_\beta = 1$) takes place. It is clear that, although for $p_\beta = 0$ the migration angle is unique ($\alpha = 0°$), that is not the case when the particles move at angles different from zero, and even for $p_\beta = 1$ there is a distribution of possible migration angles. It is therefore important to introduce a measure of separation resolution. We extend the standard definition of spatial resolution based on the position of the peaks to an angular resolution in the migration angles. Specifically, we define the resolution between species α and β as $R = \frac{(\mu_\alpha - \mu_\beta)}{2(\sigma_\alpha + \sigma_\beta)}$, where $\mu_\alpha, \mu_\beta$ are the average migration angles of the two species and $\sigma_\alpha, \sigma_\beta$ are the respective standard deviations.[39] Applying this definition for the separation of tetrahedrons and cubes at a forcing angle $\theta = 10°$ we obtain excellent resolution, $R = 1.16$. The probability distribution of tetrahedrons and cubes for a forcing angle $\theta = 10°$ are presented in Figure 6, in which the peak separation is evident. Let us note that, the same resolution corresponds to the separation between tetrahedrons and spheres or cylindrical particles, given that for $\theta \cong 10°$ they are also locked to move at $\alpha = 0°$.

It is also clear that, although still possible, the fractionation of a mixture of spheres and cubes would have a much lower resolution. For example, at a driving angle $\theta = 20°$ the average migration angle of cubes is $\alpha = 6.2°$ and $p_\beta > 0.9$, whereas the average migration angle of spherical particles is $\alpha = 1.2°$, with $p_\beta < 0.7$. In Figure 6 we present the probability distribution function for the migration angle of both cubes and spheres (2mm nominal size) and for a 20° forcing angle. In this case, we obtain a resolution $R = 0.43$, which is close to peak resolution (R=0.5).

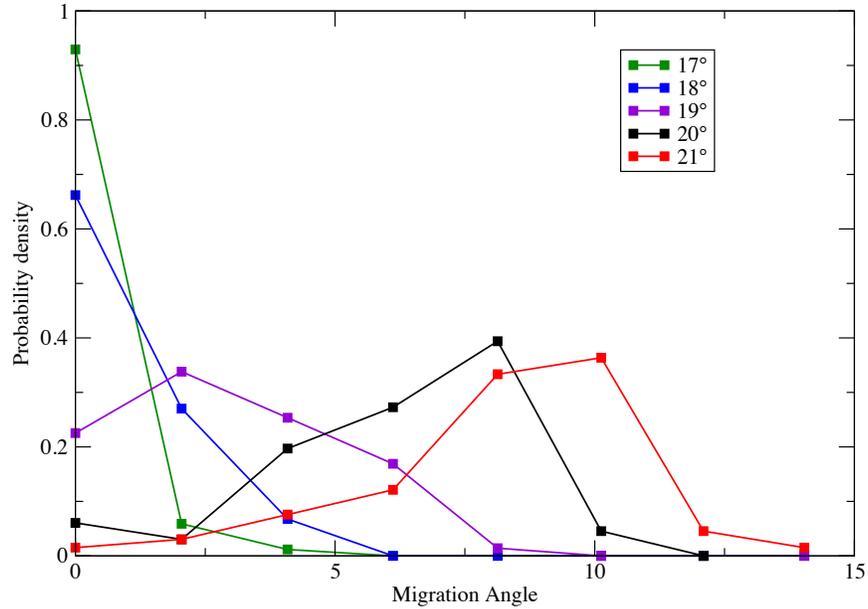

**Figure 5:** Probability distribution of migration angles for 2mm cubes. The forcing angles are indicated and correspond to those in which the probability of crossing transitions from no-crossing ($p_\beta = 0$) to complete crossing ($p_\beta = 1$).

A significant difference with previous results obtained for spherical particles, is that in the experiments discussed here the transition from no-crossing ($p_\beta = 0$) to complete-crossing ($p_\beta = 1$) happens over a finite range of forcing angles. By contrast, previous results obtained with spherical particles exhibited a sharp transition from $p_\beta = 0$ to $p_\beta = 1$ at a well-defined angle, the critical angle $\theta_c$. There are several factors that could contribute to this difference in the way particles go from locking at zero migration angle to complete crossing. First, in the case of non-spherical particles, the outcome (crossing or no-crossing) could depend on the initial orientation of the particle as it enters the array of obstacles. Therefore, intermediate values of the crossing probability would be natural. Moreover, 3D-printed spherical particles exhibit deviations from perfect spheres due to printer resolution, which could also lead to different outcomes, especially for forcing angles close to the critical angle corresponding to perfect spheres. Finally, the experiments discussed here were performed in water and there could be inertia effects, not present in previous experiments performed using higher viscosity fluids and lower Reynolds numbers. In order to investigate inertia effects in the transition from no-crossing to crossing, we performed experiments in a higher viscosity fluid, a 50% by volume mixture of glycerol and water. The results obtained for 2.5mm spherical particles are presented in Figure 7. Clearly, the transition in the crossing probability becomes significantly sharper upon increasing the viscosity of the suspending fluid. In addition, the entire probability curve shifts to lower forcing angles, consistent with previous experiments using spherical particles of different densities.[15,40,41] This also suggests the possibility to separate particles by density. Particles of different densities would also have a different Reynolds number associated with their motion in a quiescent fluid and we might expect a difference in the critical angle, thus enabling separation.

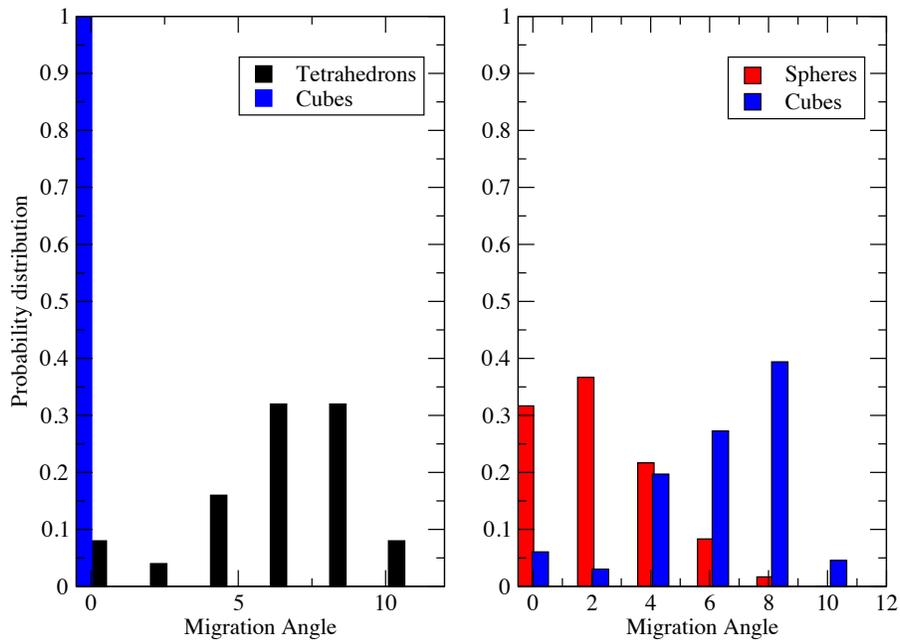

**Figure 6:** (left) Probability distribution of migration angles for 2mm cubes and tetrahedra at a forcing angle θ = 10°. (right) Probability distribution of migration angles for 2mm cubes and spheres at θ = 20°.

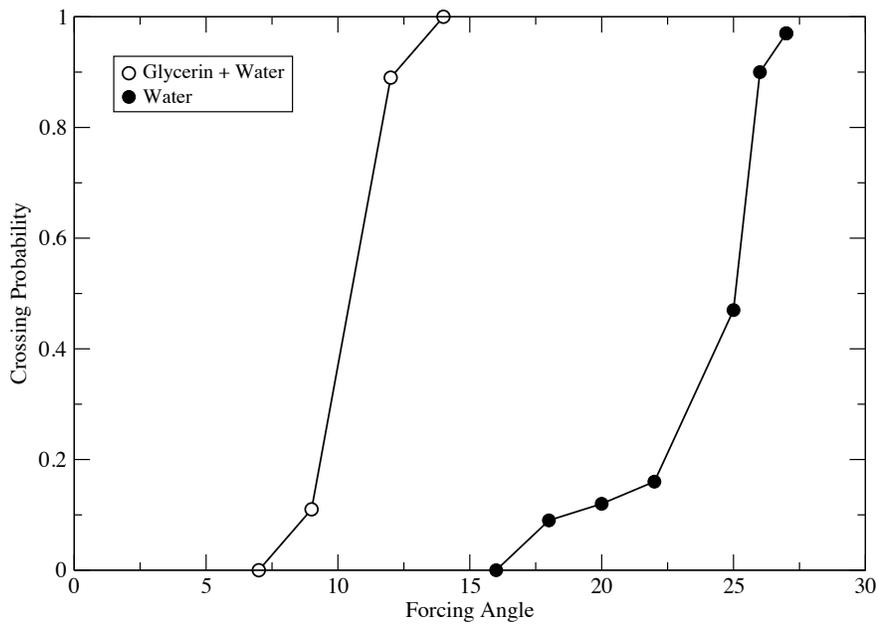

**Figure 7:** Crossing probability as a function of forcing angle for 2.5mm spherical particles in two different fluids. The open symbols correspond to a 50% by volume mixture of water and glycerin and the solid symbols correspond to water.

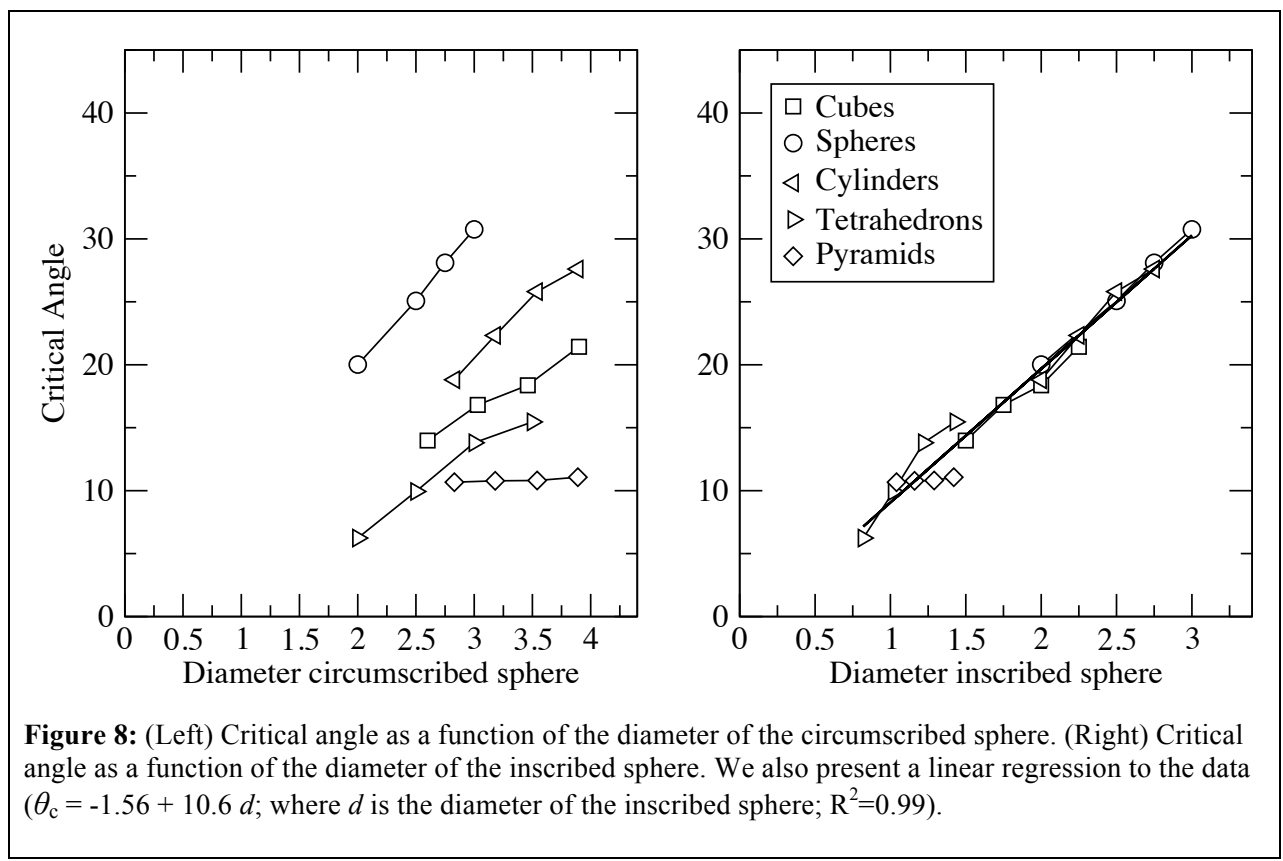

**Figure 8:** (Left) Critical angle as a function of the diameter of the circumscribed sphere. (Right) Critical angle as a function of the diameter of the inscribed sphere. We also present a linear regression to the data ($\theta_c$ = -1.56 + 10.6 $d$; where $d$ is the diameter of the inscribed sphere; $R^2$=0.99).

Finally, we investigated the existence of a characteristic length associated with the particles that could be used to predict the critical angle, independent of particle shape. The nominal size, for example, is not enough to predict the critical angle, as shown in Figure 3f. Clearly, for the same nominal size, the critical angle depends on the shape of the particle. In Figure 8 we present the critical angle as a function of both, the diameter of the circumscribed sphere on the left plot and the diameter of the inscribed sphere on the right. First, the case of the circumscribed sphere suggests that particle with the same value of circumscribed sphere but different shape will have a different critical angle, at least for those geometrical shapes considered here. More interestingly, the data seems to collapse into a unique linear relation independent of particle shape when we plot it against the diameter of the inscribed sphere. Therefore, the diameter of the inscribed sphere seems to be a good characteristic length to predict the critical angle independent of the shape of the particles. Particles with different inscribed spheres will have a different critical angle and can therefore be separated.

**Conclusions**

We have shown that gravity-driven deterministic lateral displacement (g-DLD) is a promising method for the separation of suspended particles by size and shape. We performed systematic experiments to investigate the migration of particles of five different geometrical shapes, of four different sizes each, and covering a broad range of orientations of the driving force. At relatively small forcing angles, we observe that the particles do not move across lines of obstacles (columns of obstacles) and are thus locked at a migration angle $α = 0°$. This behavior is analogous to that observed in the motion of spherical particles. However, in contrast to the case of spheres, the transition from zero migration angle at small forcing angles, to particles moving across columns of obstacles and migrating at $α \neq 0°$ as the orientation angle of the driving force increases, takes place over a range of forcing directions. We describe this transition by measuring the crossing probability $p_β$ that a particle of species $β$ would move across a column of obstacles, and define the critical angle $θ_c$ (for species $β$) as the forcing angle at which $p_β = 1/2$. We observe that the critical angle increases with particle size for all the different particles considered here, enabling size-based separation at specific force orientations. More importantly, the critical angle also depends on the geometry of the particles and shape-based separation is possible. In order to quantify the quality of separation we measured the migration angle of the particles for driving forces around each critical angle. In particular, we identify driving directions at which excellent resolution can be achieved for the separation of of binary suspensions (a mixture of tetrahedra and other particles). We also showed that, although at significantly lower resolution, it is possible to fractionate a mixture of cubes and spheres, which have a similar critical angle. Finally, we showed that using the diameter of the inscribed sphere as the characteristic size of the particles the critical angle of different particles becomes nearly independent of shape. In other words, the critical angle of particles with different diameter of the inscribed sphere is different, and it is possible to separate them. Interestingly, the dependence of the critical angle on the diameter of the inscribed sphere is linear. This linear and possibly universal behavior could provide guidance in the design and optimization of deterministic lateral displacement devices used for shape-based separation.

This work was partially supported by the National Science Foundation Grant No. CBET-1339087